\documentclass[12pt]{article}
\usepackage{amsmath}
\usepackage{amssymb}
\usepackage[titletoc]{appendix}
\begin{document}
\vskip 2cm
\begin{center}
{\sf {\Large   $\mathcal{N }= 2$ SUSY symmetries  for  a moving  charged particle 
under influence of a  magnetic field: Supervariable approach}}

\vskip 3.0cm

{\sf S. Krishna$^{(a)}$, R. P. Malik$^{(a,b)}$}\\
$^{(a)}$ {\it Physics Department, Centre of Advanced Studies,}\\
{\it Banaras Hindu University, Varanasi - 221 005, (U.P.), India}\\
$^{(b)}$ {\it DST Centre for Interdisciplinary Mathematical Sciences,}\\
{\it Faculty of Science, Banaras Hindu University, Varanasi - 221 005, India}\\
{\small {\sf {e-mails: skrishna.bhu@gmail.com;  rpmalik1995@gmail.com}}}

\end{center}

\vskip 2cm

\noindent
{\bf Abstract:} 
We exploit the supersymmetric invariant restrictions (SUSYIRs) on the supervariables to derive the 
nilpotent $\mathcal{N} = 2$ SUSY transformations for the supersymmetric quantum mechanical model
of the motion of a charged particle in the X-Y plane (where the magnetic field ($B_z$) is applied along the
Z-direction). The supervariables are defined on a (1, 2)-dimensional supermanifold parametrized   
by a bosonic ``time" variable $t$ and a pair of Grassmannian variables $\theta$ and $\bar\theta$
(with $\theta^2 =  {\bar\theta}^2 = 0,\; \theta\bar\theta + \bar\theta\theta = 0$). 
We take the (anti-)chiral supervariables for our purpose so that the nilpotency property 
of the  $\mathcal{N} = 2$ SUSY  symmetry transformations could be captured within the framework of 
supervariable approach. We express the Lagrangian as well as supercharges in terms of the  
supervariables (that are obtained after the application of the appropriate SUSYIRs)
and provide geometrical basis, within the framework of our supervariable approach, 
for ($i$) the nilpotency property of the above SUSY transformations (and the corresponding supercharges),
and ($ii$) the SUSY invariance of the Lagrangian.

\vskip 0.8cm
\noindent
PACS numbers:  11.30.Pb, 03.65.-w, 02.40.-k

\vskip 0.5cm
\noindent
{\it Keywords}:  Supervariable approach;   $\mathcal{N }= 2$ SUSY quantum mechanical model;  
 SUSY transformations; nilpotency property; (anti-)chiral supervariables; geometrical basis

\newpage
\section{Introduction}

The well-known Becchi-Rouet-Stora-Tyutin (BRST) formalism is one of the mathematically rich 
and theoretically useful approaches to covariantly quantize the gauge  theories where 
the local gauge symmetries of the original theory are traded with the nilpotent (anti-)BRST symmetries. 
Two of the  abstract mathematical properties associated with the above (anti-)BRST symmetries   
are the nilpotency and absolute anticommutativity. 
The geometrical origin and  interpretations  for the above mentioned 
nilpotency and anticommutativity  
properties are provided  by  the superfield formalism 
[1-8]. In particular, the Bonora-Tonin (BT)
superfield approach [4,5]  has been very successful    
in the context of gauge theories  where  the horizontality  (HC) condition plays a very crucial role. 
The latter condition leads to the derivation  of ``quantum" (anti-)BRST symmetries  for the gauge and corresponding 
(anti-)ghost fields which turn out to be nilpotent of order two and absolutely anticommuting
{\it but} it does {\it not} say anything about matter fields. 


In a set of papers [9-13], we have extended the BT-superfield 
formalism\footnote{This extended version of the geometrical BT-superfield formalism 
has been christened as the augmented 
version of superfield formalism [9-13].} where, 
in addition to the HC, we have exploited  the gauge  invariant restrictions (GIRs) to 
derive the (anti-)BRST symmetry transformations  for the {\it matter} fields, too, in an {\it interacting} 
gauge theory. The symmetries (and their geometrical interpretations) turn out to be consistent with one-another
when the HC and GIRs are tapped {\it together} within the framework of  augmented  version
of BT superfield formalism [9-13]. It has been a long-standing problem to apply the above 
superfield formalism [1-13] to derive the SUSY transformations for the SUSY systems where the nilpotency  
property exists  but the anticommuting property does {\it not}. In a very recent set of 
papers [14,15], however,  we have 
applied the augmented version of superfield/supervariable formalism [9-13]  to derive the $\mathcal{N} = 2$ SUSY 
transformations in a consistent and cogent  manner (for the specific $\mathcal{N} = 2$ 
SUSY quantum mechanical models). We have coined the word {\it supervariable approach} for our method
of derivation of SUSY symmetries (cf. footnote just before (10)).

The supervariable approach [14,15] to derive the SUSY symmetries for the $\mathcal{N} = 2$  quantum mechanical
systems is a {\it novel} approach in the context of SUSY theories.
The purpose of our present investigation is to exploit  the theoretical tools and techniques of our earlier 
works  on the supervariable approach  [14,15] to derive the $\mathcal{N} = 2$   SUSY 
transformations for the SUSY system  of a moving charged particle in the X-Y plane 
under influence of a magnetic field 
that is applied along the Z-direction 
(i.e. perpendicular  to the X-Y plane). We express the conserved charges and   Lagrangian in the language of
supervariables  and provide the geometrical interpretations for the SUSY invariance of the Lagrangian
as well as the nilpotency  of the conserved charges in terms of the translational generators along the 
Grassmannian directions ($\bar\theta)\theta$ of the (anti-)chiral super-submanifolds, 
respectively. These generators  are defined on the (1, 1)-dimensional super-submanifolds of 
the general (1, 2)-dimensional supermanifold
on which our {\it starting} theory 
is  generalized within the framework of supervariable approach.


Our present investigation has been motivated by the following key factors.
First and foremost, to put   our central ideas [14,15] on a solid-footing, 
it is essential that we should  apply the supervariable  approach to the models with superpotentials that
 are completely  different from  the  superpotentials of the  $\mathcal{N} = 2$ SUSY free 
particle, harmonic oscillator (HO) and the generalized version of the HO [16,17]. 
This is the reason  that, in our present investigation, 
we have taken the SUSY example of the motion of a charged particle  under  influence of a magnetic 
field
and have demonstrated  the utility of our supervariable
approach. Second, it has been a long-standing problem to apply some {\it form} of the
superfield approach [1-13] to capture the nilpotency of the SUSY symmetries  and provide a geometrical meaning to it.
We have accomplished  this  goal in our present investigation  (and in our earlier works
[14,15]). Finally, our method of application of supervariable/superfield  formalism might  turn out 
to be useful in the context of SUSY gauge theories. 

The contents of our present investigation are organized  as follows.
First of all, we discuss the bare  essentials of  the $\mathcal{N} = 2$ SUSY transformations 
for the motion of a charged particle under  influence of a magnetic field in Sec. 2. 
We exploit the virtues  of (anti-)chiral supervariables to derive  the two
nilpotent $\mathcal{N} = 2$ SUSY transformations in Sec. 3. 
We discuss about the SUSY invariance of the 
Lagrangian of the theory and nilpotency of the $\mathcal{N} = 2$ SUSY charges within the framework
of supervariable  formalism, in our Sec. 4. Our Sec. 5 deals with the cohomological 
aspects  of the $\mathcal{N} = 2$
SUSY transformations and corresponding symmetry generators. Finally, 
in Sec. 6, we make some concluding remarks
and point out a few future directions for further investigation.

We provide the logical reasons behind our choice of the (anti-)chiral 
supervariables in our Appendix A.

\section{Preliminaries: $\mathcal{N} = 2$ SUSY symmetries}

We begin with the following  Lagrangian for the motion of a  charged particle 
(of mass $ m = 1$ and charge $ e = 1$)
in the X-Y plane  (see, e.g. [16,17]):  
\begin{eqnarray}
L_0 = \frac{1}{2}\,(\dot x^2 + \dot y^2)- (\dot x\, A_x + \dot y \, A_y) + i\, \bar\psi\,\dot\psi 
+ B_z\, \bar\psi\, \psi,
\end{eqnarray}
where the magnetic field $B_z = \partial_x\, A_y (x,y) - \partial_y\, A_x (x,y)$ is in the Z-direction 
and the whole trajectory  of the particle is parametrized by the evolution ``time"  parameter $t$.
As a consequence, we have the ``generalized" instantaneous velocities of the SUSY particle as: 
 $\dot x = dx/dt,\, \dot y = dy/dt$ and $\dot\psi = d\psi/dt$. The instantaneous position variables 
$x(t)$ and $y(t)$ are bosonic in nature and variables $\psi(t)$ and $\bar\psi(t)$ are fermionic 
($\psi^2 = {\bar\psi}^2 = 0,\, \psi\,\bar\psi +\bar\psi\, \psi = 0$) at the {\it classical}
level. The X-Y components of the vector potentials ($A_x, A_y$) have no {\it explicit}
 ``time" dependence and they are {\it only} function of the
instantaneous position of the particle (i.e. $A_x (x, y),\, A_y (x, y)$).  

It can be readily checked that the starting Lagrangian (1)  respects the following $\mathcal{N} = 2$
SUSY transformations   $s_1$ and $s_2$ [16,17]   
\begin{eqnarray}
&& s_1 x = {\psi}, \;\;\;\quad s_1 y = {-i\,\psi},\; \;\;\quad
  s_1 \bar \psi = i\, (\dot x - i \,\dot y), 
\;\;\;\quad s_1 \psi = 0, \nonumber\\ &&  s_1 A_x =  \bigl (\partial_x A_x 
- i\, \partial_y A_x \bigr ) \, \psi, \;\;\;\qquad
s_1 A_y =  \bigl (\partial_x A_y - i \,\partial_y A_y \bigr )\, \psi, 
 \nonumber\\ &&  s_2 x = {\bar \psi}, \qquad s_2 y = {i\,\bar \psi}, \qquad 
 s_2 \bar \psi = 0, \qquad    s_2  \psi = i\, (\dot x + i \,\dot y), \nonumber\\ && s_2 A_x 
= \bar \psi \,\bigl (\partial_x A_x + i \,\partial_y A_x \bigr ), \;\;\;\qquad
 s_2 A_y = \bar \psi\, \bigl (\partial_x A_y + i \,\partial_y A_y \bigr ), 
\end{eqnarray}
because the Lagrangian transforms as:
\begin{eqnarray}
&& s_1 L_0 = \,- \,\frac{d}{d t} \,\Bigl [(A_x - i\, A_y) \,\psi  \Bigr ], \nonumber\\
&&s_2 L_0 = \,+\, \frac{d}{d t} \,\Bigl [\bar \psi \,
\bigl \{\dot x + i\, \dot y - (A_x + i\, A_y) \bigr \} \Bigr ].
\end{eqnarray}  
This establishes   that the relevant action integral $S = \int dt\, L_0$ remains invariant 
under the continuous transformations  $s_1$ and $s_2$.

We point out that the above infinitesimal transformations are off-shell nilpotent 
of order two (i.e. $s^2_1 = s_2^2 = 0$) which  establishes their  fermionic nature.  
This is the reason that the above transformations
 change bosonic variables into fermionic
 variables and {\it vice-versa}. Furthermore, we note that the anticommutator  
of the fermionic transformations $s_1$ and
$s_2$ leads to a bosonic symmetry transformation (i.e. $s_\omega = \{s_1,\, s_2\}$), namely; 
\begin{eqnarray}
&& s_\omega \; \Phi =  \; \dot \Phi, \qquad  \Phi = x (t),\; y (t), \;\psi (t),\; \bar \psi (t),\;
 A_x (x, y),\; A_y (x, y),
\end{eqnarray}
modulo a factor of ($2\,i$). In the derivation of the above bosonic symmetry transformations, we have used
(for obvious reasons) the following inputs:
\begin{eqnarray}
&& \partial_x \psi (t) = 0, \quad\qquad \partial_y \psi (t) = 0, \quad\qquad 
\partial_x \bar \psi (t) = 0,  \quad\qquad
 \partial_y \bar\psi (t) = 0,\nonumber\\ && \frac{d}{dt} A_x (x, y) 
= \dot x \; \partial_x A_x + \dot y\; \partial_y A_x, \qquad
\frac{d}{dt} A_y (x, y) = \dot x \; \partial_x A_y + \dot y\; \partial_y A_y.
\end{eqnarray}
Under the above transformations (4), the starting Lagrangian $L_0$ transforms to a total 
``time" derivative (of itself) as follows:
\begin{eqnarray}
s_\omega \; L_0  = \frac{d}{dt} \left [ L_0 \right ], 
\end{eqnarray} 
which demonstrates the invariance of the action integral $S = \int dt \,L_0$.
It is straightforward 
to check that $s_\omega$ commutes with {\it both} the fermionic  transformations $s_{(1)2}$
(i.e. $[s_\omega,\, s_1] = 0, [s_\omega,\, s_2] = 0$).

According  to Noether's theorem, the above continuous  transformations 
lead to the derivation  of conserved charges $Q_i$ (with $ i = 1, 2, 3$) as 
\begin{eqnarray}
&& Q_1 \equiv Q =  \Bigl [ (p_x + A_x) - i \;(p_y + A_y) \Bigr ] \; \psi, \quad\qquad p_x = \dot x, \nonumber\\
&& Q_2 \equiv \bar Q = \bar\psi\,\Bigl [ (p_x + A_x) + i \;(p_y + A_y) \Bigr ],\quad\;\qquad p_y = \dot y, \nonumber\\
&& Q_3 \equiv Q_\omega  =  \left [\frac{(p_x + A_x)^2}{2} +  \frac{(p_y + A_y)^2}{2} 
- B_z\; \bar\psi\; \psi \right ] \; \equiv H.
\end{eqnarray}
The conservation  ($ \dot Q_i = 0$) of the above charges  
$Q_i$ can be proven   directly  by using the following
Euler-Lagrange equations of motion:
\begin{eqnarray}
\dot\psi - i\, B_z\, \psi = 0, 
\qquad \ddot x + \dot y\, B_z\, - (\partial_x B_z)\, \bar\psi\,\psi =0, \nonumber\\ 
 \dot{\bar\psi} + i\, B_z\, \bar\psi = 0,\qquad
\ddot y - \dot x\, B_z\, -  (\partial_y B_z)\, \bar\psi\,\psi =0,
\end{eqnarray}
which are derived  from the Lagrangian $L_0$. The above conserved charges   are the generators of the 
infinitesimal symmetry  transformations listed in (2) and (4). This can be explicitly 
checked by the following general relationship
for the generic variable $\Phi$ of our present theory, namely; 
\begin{eqnarray}
s_r \, \Phi = \pm \,i\, [\Phi,\, Q_r]_{\pm},\qquad\qquad r = 1,\, 2,\, \omega, 
\end{eqnarray} 
where the ($\pm$) signs (expressed as the subscripts)  on the square bracket correspond to the (anti)commutator
for the generic variable $\Phi =  x, y, A_x, A_y, \psi, \bar\psi$
being (fermionic)bosonic in nature. 

\section{(Anti-)chiral supervariables: SUSY transformations}

To derive the transformations $s_1$ and its
nilpotency, we choose the anti-chiral supervariables 
(corresponding to {\it all} the ordinary dynamical variables of the starting Lagrangian $L_0$) 
on the (1, 1)-dimensional  super-submanifold of the general (1, 2)-dimensional supermanifold
 on which our present SUSY theory is generalized\footnote{We are theoretically compelled  to choose 
the (anti-)chiral supervariables because the nilpotent  $\mathcal{N} = 2$ SUSY transformations 
do {\it not} anticommute (i.e. $\{s_1,\, s_2\}\ne 0$).
This should be contrasted  with the nilpotent (anti-)BRST symmetry transformations which 
absolutely anticommute (see, e.g. [9-13] for details). Within the framework of superfield approach 
to  (anti-)BRST symmetries, the superfields are expanded  along both the Grassmannian 
directions ($\theta, \bar\theta$)  (see, e.g. Appendix  A).}.  
In other words, first of all, we generalize the simple variables ($x(t), y(t),  \psi(t), 
\bar\psi (t)$) onto the (1, 1)-dimensional anti-chiral super-submanifold as 
anti-chiral supervariables\footnote{We observe that, in the limit $\bar\theta = 0$, 
we get back the  {\it variables} $x(t), y(t), \psi(t)$ and $\bar\psi(t)$ 
from (10). This is why we have christened our present 
technique as the {\it supervariable approach} to the description of some $\mathcal{N} = 2$ 
SUSY quantum mechanical models.} (see, e.g. [14,15]): 
\begin{eqnarray}
&&x(t)  \longrightarrow  X(t, \theta, \bar\theta)\mid_{\theta = 0}  
\equiv X(t,\bar\theta) = x(t) +  \bar\theta\, f_1(t),\nonumber\\
&&y(t)  \longrightarrow  Y(t, \theta, \bar\theta)\mid_{\theta = 0} 
\equiv Y(t,\bar\theta) = y(t) +  \bar\theta\, f_2(t),\nonumber\\
&&\psi(t) \longrightarrow  \Psi (t, \theta, \bar\theta) \mid_{\theta = 0}
\equiv \Psi (t,  \bar\theta) = \psi (t)  + i\, \bar\theta\, b_1 (t), \nonumber\\
&&\bar\psi(t) \longrightarrow  \bar\Psi (t, \theta, \bar\theta)\mid_{\theta = 0} 
\equiv \bar\Psi (t,  \bar\theta) = \bar\psi (t)  + i\, \bar\theta \,b_2 (t),
\end{eqnarray} 
where  the secondary variables ($b_1, b_2$) and ($f_1, f_2$) are bosonic and fermionic 
in nature, respectively. We  note that the bosonic 
(i.e. $x, y, b_1, b_2)$ and fermionic ($\psi, \bar\psi, f_1, f_2$) d.o.f.
{\it do} match on the r.h.s.  of the above anti-chiral  expansions (cf.  (10)) which is one of 
the key requirements of a SUSY theory.

A decisive feature of the augmented version of BT-superfield formalism [9-13] and 
our earlier works [14,15] is the requirement that {\it all} the  gauge/SUSY  invariant 
quantities must remain independent of the Grassmannian
variables $\theta$ and $\bar\theta$ when they are generalized onto a specific supermanifold.
We observe that such invariant quantities,  w.r.t. $s_1$, are as follows:
\begin{eqnarray}
&& s_1\,[\psi(t)] = 0,\quad s_1[x(t)\,\psi(t)] = 0,
 \quad s_1[y(t)\,\psi(t)] = 0, \quad
 s_1 \,[\dot{x}(t) \,\dot\psi(t)] = 0, \nonumber\\ &&
s_1 \,[\dot{y}(t) \,\dot\psi(t)] = 0,
\qquad\qquad s_1\, \left[\frac{1}{2}\,\left(\dot x^2 (t)  + \dot y^2(t) \right) 
 + i\, \bar\psi(t)\,\dot\psi(t) \right] = 0.
\end{eqnarray}
As per prescription laid down in [14,15], we have the following SUSY invariant restrictions (SUSYIRs)
on the (super)variables: 
\begin{eqnarray}
&& X(t, \bar\theta)\,\Psi (t, \bar\theta) = x(t)\, \psi (t),\qquad
 Y(t, \bar\theta)\,\Psi (t, \bar\theta) = y(t)\, \psi (t),\nonumber\\ &&
{\dot X}(t, \bar\theta)\,\dot\Psi (t, \bar\theta) = {\dot x}(t)\,  \dot\psi (t),\qquad
{\dot Y}(t, \bar\theta)\,\dot\Psi (t, \bar\theta) = {\dot y}(t)\, \dot\psi (t), \nonumber\\
&&  \frac{1}{2}\left[\dot {X}^2(t, \bar\theta) + \dot {Y}^2(t, \bar\theta) \right] 
+ i\, \bar\Psi (t, \bar\theta)\, \dot{\Psi}(t, \bar\theta) 
= \frac{1}{2} \,\Big[\dot x^2(t)   + \dot y^2(t)\Big]  \nonumber\\ && + i\, \bar\psi (t)\, \dot\psi (t),
\qquad \Psi (t, \bar\theta) = \psi (t),
\end{eqnarray}
which lead to the following relationships  amongst the secondary  variables ($b_1, b_2, f_1, f_2$)
of the expansions (10) and the basic variables $(x, y, \psi, \bar\psi)$, namely;
\begin{eqnarray}
&& b_1(t) = 0, \,\qquad f_1(t)\,\psi (t) = 0,\,\qquad \dot f_1(t)\,\dot\psi (t) = 0, \qquad
f_2(t)\,\psi (t) = 0, \nonumber\\ &&  \dot f_2(t)\,\dot\psi (t) = 0, \qquad\qquad
\dot x(t) \,{\dot f}_1 (t)+ \dot y (t)\,{\dot f}_2(t) - b_2(t) \,\dot\psi (t) = 0.
\end{eqnarray}
The non-trivial solution of the above restrictions are  $f_1(t) \propto \psi (t)$ and 
 $f_2 (t) \propto \psi (t)$. 
For the algebraic convenience, however,  we choose $f_1(t) = \psi(t)$ and $f_2(t) = - i\,\psi(t)$. It is evident
that if we take the help of these relationships, we obtain $b_2 = \dot x - i \dot y$
from the last entry of (13).

The explicit  substitution of ($f_1, f_2, b_1, b_2$) into the original expansions (10) 
leads to the following {\it final} expansions
of the anti-chiral supervariables  
\begin{eqnarray}
&& X^{(1)}(t, \bar\theta) = x(t) +  \bar\theta\, (\psi) \equiv x(t) + \bar\theta \,(s_1\, x),\nonumber\\
&& Y^{(1)}(t, \bar\theta) = y(t) +  \bar\theta\, (-i\,\psi) \equiv y(t) + \bar\theta \,(s_1\, y),\nonumber\\
&& \Psi^{(1)} (t, \bar\theta) = \psi (t)  + \bar\theta\,(0) \equiv \psi(t) + \bar\theta\, (s_1\, \psi), \nonumber\\
&& \bar\Psi^{(1)} (t, \bar\theta) = \bar\psi (t)  
+  \bar\theta\,\left[i\, (\dot x - i  \dot y)\right] 
\equiv \bar\psi (t)  + \, \bar\theta\, (s_1  \bar\psi),
\end{eqnarray} 
where the superscript $(1)$, placed  on the supervariables, denotes the expansions of the
supervariables after the application of the SUSYIRs (12). It is evident now that
the following geometrical relationship between the SUSY transformations $s_1$
and the translational generators $\partial_{\bar\theta}$ 
emerges in an explicit fashion [cf. equation (9)]:
\begin{eqnarray}
\frac{\partial}{\partial\bar\theta}  \left[\Omega^{(1)} (t, \theta, \bar\theta)\right]|_{\theta = 0}
 = s_1 \, \Omega (t)  \equiv \pm\, i \, [\Omega(t), \, Q]_\pm,
\end{eqnarray}
where $\Omega (t)\equiv x(t), y(t), \psi (t), \bar\psi(t)$  is the generic variable 
of the starting Lagrangian $L_0$  and $\Omega^{(1)} (t, \theta, \bar\theta)|_{\theta = 0}$ 
stands for the generic supervariables (14) that 
have been obtained after application of the SUSYIRs (12). A close and careful look at (15) and (14)
explains clearly that we have already obtained the SUSY transformations\footnote{It will be noted that our supervariable approach allows us to choose the secondary variables as has been done in (14) {\it 
modulo  a constant factor}. This freedom would be exploited in our Sec. 5 for some specific purpose.}: 
$s_1 x = {\psi}, \, s_1 y = {-i\,\psi}, \,
s_1 \psi = 0, \, s_1 \bar \psi = i\, (\dot x - i \,\dot y)$ which are present in (2).
Their nilpotency is also clear because of the relationship in (15) which states that $s^2_1  = 0$ and $(\partial_{\bar\theta})^2 = 0$ are inter-related.

Let us now  focus on the SUSY transformations for $A_x$ and $A_y$ and
point out  the derivation of $s_1 A_x$ and $s_1 A_y$ within the 
framework of our supervariable approach. Towards  this goal in mind, first of all, we generalize the 
ordinary potentials $A_x(x,y)$ and $A_y (x, y)$ onto their counterpart  
anti-chiral supervariables on the anti-chiral  super-submanifold ($x\rightarrow  X^{(1)},\, 
y \rightarrow  Y^{(1)},\, A_x \rightarrow  {\tilde A}_x,\, A_y \rightarrow  {\tilde A}_y$) as
\begin{eqnarray}
&& A_x(x, y) \longrightarrow  {\tilde A}_x (X^{(1)},\, Y^{(1)}) 
\equiv {\tilde A}_x (x + \bar\theta \,\psi,\, y - i\,\bar\theta \,\psi)\nonumber\\
&& ~~~~~~~= A_x (x, y) + \bar\theta\, \Big[\bigl(\partial_x A_x (x, y) 
- i\,\partial_y A_x (x, y)\bigr)\,\psi\Big] \nonumber\\
&&~~~~~~~\equiv  A_x (x,y) + \bar\theta \,\bigl(s_1\, A_x (x, y)\bigr), \nonumber\\
&& A_y(x, y) \longrightarrow {\tilde A}_y (X^{(1)},\, Y^{(1)}) \equiv {\tilde A}_y (x + \bar\theta \,\psi,\, y 
- i\,\bar\theta \,\psi)\nonumber\\
&&~~~~~~~= A_y (x, y) + \bar\theta\, \Big[\bigl(\partial_x A_y (x, y) 
- i\,\partial_y A_y (x, y) \bigr)\,\psi \Big] \nonumber\\
&&~~~~~~~ \equiv  A_y (x,y) + \bar\theta \,\bigl(s_1\, A_y (x, y)\bigr).
\end{eqnarray}
We note that we have to use the expansions, obtained in (14), for the derivation of  SUSY transformations
$s_1\, A_x(x, y)$ and $s_1 \,A_y(x, y)$ which are
\begin{eqnarray}
s_1 A_x =  \bigl (\partial_x A_x - i\, \partial_y A_x \bigr ) \, \psi, \qquad
s_1 A_y =  \bigl (\partial_x A_y - i \,\partial_y A_y \bigr )\, \psi. 
\end{eqnarray}
It is clear  that our above results match with the ones listed in (2). 


To derive the other SUSY transformations $s_2$, beside the first one (i.e. $s_1$),
we take recourse  to the chiral supervariables that are generalization of, first of all, the
{\it simple} dynamical variables 
($x(t), y(t), \psi(t), \bar\psi(t)$) of the starting Lagrangian $L_0$.
In other words, we generalize the ordinary SUSY theory onto a (1, 1)-dimensional chiral super-submanifold 
 as [14,15]
\begin{eqnarray}
&&x(t)  \longrightarrow  X(t, \theta, \bar\theta)\mid_{\bar\theta = 0} 
\equiv X(t,\theta) = x(t) +  \theta\, {\bar f}_1(t),\nonumber\\
&&y(t)  \longrightarrow  Y(t, \theta, \bar\theta)\mid_{\bar\theta = 0}
\equiv Y(t, \theta) = y(t) +  \theta\, {\bar f}_2(t),\nonumber\\
&&\psi(t) \longrightarrow  \Psi (t, \theta, \bar\theta) \mid_{\bar\theta = 0}
\equiv \Psi (t,  \theta) = \psi (t)  + i\, \theta\, {\bar b}_1 (t), \nonumber\\
&&\bar\psi(t) \longrightarrow  \bar\Psi (t, \theta, \bar\theta)\mid_{\bar\theta = 0} 
\equiv \bar\Psi (t,  \theta) = \bar\psi (t)  + i\, \theta\, {\bar b}_2 (t),
\end{eqnarray} 
where   (${\bar b}_1, {\bar b}_2$) and (${\bar f}_1, {\bar f}_2$)
 are the bosonic and fermionic  secondary variables, respectively. It is crystal clear that the 
bosonic ($x, y, {\bar b}_1, {\bar b}_2 $) and fermionic ($\psi, \bar\psi, {\bar f}_1, {\bar f}_2 $)
d.o.f. {\it do} match on the r.h.s. of the expansions (18) which is a key requirement of 
any arbitrary SUSY theory.

As proposed in the augmented version of superfield formalism [9-13]
and in our earlier works [14,15], we have to find out the SUSY invariant quantities under $s_2$
and demand that they should be independent of the Grassmannian variables $\theta$ and $\bar\theta$
when they are generalized onto the appropriate  supermanifold. In this regards, we note that 
the following 
\begin{eqnarray}
&& s_2\,[\bar\psi(t)] = 0,\quad s_2[x(t)\,\bar\psi(t)] = 0, 
\quad s_2[y(t)\,\bar\psi(t)] = 0, \quad s_2 \,[\dot{x}(t) \,\dot{\bar\psi}(t)] = 0,\nonumber\\ &&
  s_2 \,[\dot{y}(t) \,\dot{\bar\psi}(t)] = 0, \qquad
s_2\, \left[\frac{1}{2}\,\left(\dot x^2 (t) + \dot y^2(t) \right)  - i\, \dot{\bar\psi} (t)\,\psi(t) \right] = 0.
\end{eqnarray}
As a consequence, we have the following  interesting and important SUSYIRs  
on the (super)variables, namely; 
\begin{eqnarray}
&& X(t, \theta)\,\bar\Psi (t, \theta) = x(t)\, \bar\psi (t),\qquad 
Y(t, \theta)\,\bar\Psi (t, \theta) = y(t)\, \bar\psi (t),\nonumber\\
&& {\dot X}(t, \theta)\,\dot{\bar\Psi} (t, \theta) = {\dot x}(t)\, \dot{\bar\psi} (t),\qquad
{\dot Y}(t, \theta)\,\dot{\bar\Psi} (t, \theta) = {\dot y}(t)\, \dot{\bar\psi} (t), \nonumber\\
&& \frac{1}{2}\,\Big[\dot {X}^2(t, \theta) + \dot {Y}^2(t, \theta) \Big] 
- i\, \dot{\bar\Psi} (t, \theta)\, {\Psi}(t, \theta) 
= \frac{1}{2} \,\Big[\dot x^2(t)  + \dot y^2(t) \Big]  \nonumber\\ && - i\, \dot{\bar\psi} (t)\, \psi (t), \quad
 \qquad \bar\Psi (t, \theta) = \bar\psi (t).
\end{eqnarray}
Using the expansions  from (18), we obtain the following:
\begin{eqnarray}
&& {\bar b}_1(t) = 0, \qquad {\bar f}_1(t)\,\bar\psi (t) = 0,\qquad 
\dot {\bar f}_1(t)\,\dot{\bar\psi} (t) = 0,  \qquad
 {\bar f}_2(t)\,\bar\psi (t) = 0, \nonumber\\ &&  \dot {\bar f}_2(t)\,\dot{\bar\psi} (t) = 0, \qquad\qquad
 \dot x (t)\, \dot {\bar f}_1 (t)+ \dot y(t) \,\dot {\bar f}_2(t) - {\bar b}_1(t) \,\dot{\bar\psi} (t) = 0.
\end{eqnarray}
The non-trivial solution of the above restrictions are  ${\bar f}_1(t) \propto \bar\psi (t)$ and 
 ${\bar f}_2 (t) \propto \bar\psi (t)$. 
For the algebraic convenience, however, we choose ${\bar f}_1(t) = \bar\psi(t)$ and 
${\bar f}_2(t) =  i\,\bar\psi(t)$. Using these values (i.e. ${\bar f}_1 = \bar\psi,\,
{\bar f}_2 =  i\,\bar\psi$), we obtain ${\bar b}_1 =  \dot x + i\, \dot y$.
The substitution of the above secondary variables in the equation (18) of the supervariable expansions
leads to 
\begin{eqnarray}
&& X^{(2)}(t, \theta) = x(t) +  \theta\, (\bar\psi) \equiv x(t) + \theta \,(s_2\, x),\nonumber\\
&& Y^{(2)}(t, \theta) = y(t) +  \theta\, ( i\,\bar\psi) \equiv y(t) + \theta \,(s_2\, y),\nonumber\\
&& \bar\Psi^{(2)} (t, \theta) = \bar\psi (t)  + \theta\,(0) \equiv \bar\psi(t) + \theta\, (s_2\, \bar\psi), \nonumber\\
&& \Psi^{(2)} (t, \theta) = \psi (t)  
+  \theta\,\left[i\, (\dot x + i \dot y)\right] 
\equiv \psi (t)  +  \theta\, (s_2 \psi),
\end{eqnarray} 
where the superscript (2) stands for the expansions  obtained after the application of SUSYIRs (20).
It is clear, from the above expansions, that we have already derived the nilpotent transformations $s_2$
(cf. (2)).

We  derive 
the SUSY transformations $s_2$ for the potential functions 
$A_x (x, y)$ and $A_y (x, y)$. First of all, we generalize these ordinary variables onto the 
(1, 1)-dimensional chiral  super-submanifold as follows [cf. (22)]:
\begin{eqnarray}
&& A_x(x, y) \longrightarrow  {\tilde A}_x (X^{(2)}, \,Y^{(2)}) 
\equiv {\tilde A}_x (x + \theta \,\bar\psi,\, y + i\,\theta \,\bar\psi)\nonumber\\
&& ~~~~~~= A_x (x, y) + \theta\, \Big[ \bar\psi\, \bigl(\partial_x A_x (x, y) 
+ i\,\partial_y A_x (x, y)\bigr)\Big] \nonumber\\
&&~~~~~~ \equiv  A_x (x,y) + \theta \,\bigl(s_2\, A_x (x, y)\bigr), \nonumber\\
&& A_y(x, y) \longrightarrow  {\tilde A}_y (X^{(2)},\, Y^{(2)}) \equiv {\tilde A}_y (x 
+ \theta\, \bar \psi,\, y + i\, \theta \,\bar \psi)\nonumber\\
&&~~~~~~ = A_y (x, y) + \theta\, \Big[\bar\psi \,\bigl(\partial_x A_y (x, y) 
+ i\,\partial_y A_y (x, y) \bigr)\Big] \nonumber\\
&&~~~~~~\equiv  A_y (x,y) + \theta \,\bigl(s_2\, A_y (x, y)\bigr).
\end{eqnarray}
A careful observation at the expansions (22) and (23) demonstrates  that we have already obtained the SUSY
transformations $s_2$ [cf. (2)] for all  the relevant variables  of the theory\footnote{We would 
like to emphasize that all our transformations can be modified by a constant factor 
without violating the sanctity of our method. We have used such kind of modifications in our Sec. 5
for some specific purposes.}. 
We further note that 
the following mappings do exist, namely;
\begin{eqnarray}
\frac{\partial}{\partial\theta}  \left[\Sigma^{(2)} (t, \theta, \bar\theta)\right]|_{\bar\theta = 0}
 = s_2 \, \Sigma (t) \equiv \pm\, i \, [\Sigma(t), \,\bar Q]_\pm,
\end{eqnarray}
where $\Sigma (t)$ is the generic variable of the Lagrangian (1)  
[i.e. $\Sigma (t)= x(t), y(t)$, $\psi (t), \bar\psi(t)$, $A_x (x, y), A_y (x, y)$] 
and $\Sigma^{(2)} (t, \theta, \bar\theta)|_{\bar\theta = 0}$
denotes the supervariables [cf. (22), (23)] that have been obtained after the application 
of the SUSYIRs (20).  We  note  (from (24)) that the nilpotent symmetry transformations
 $s_2$ and corresponding charge $\bar Q$
are intimately related to the translational generator $\partial_\theta$ along 
the Grassmannian direction  of the chiral super-submanifold.

\section{Invariance of Lagrangian and nilpotency of supercharges: Geometrical supervariable approach }

As far as the invariance of the Lagrangian $L_0$ of (1), under the SUSY symmetry transformations 
$s_1$ is concerned, we observe that the starting Lagrangian $L_0$ can be generalized onto a (1, 1)-dimensional 
{\it anti-chiral} supermanifold in the following manner:
\begin{eqnarray}
 L_0  &\Rightarrow &     {\tilde L}^{(ac)}_0 = \frac{1}{2}\,\Big[{\dot X}^{(1)}(t, \bar\theta)\,{\dot X}^{(1)}(t, \bar\theta) + {\dot Y}^{(1)}(t, \bar\theta)\,{\dot Y}^{(1)}(t, \bar\theta)\Big] 
 \nonumber\\ 
&-& \Big[{\dot X}^{(1)}(t, \bar\theta) \, {\tilde A}_x(X^{(1)}, Y^{(1)})  +
{\dot Y}^{(1)}(t, \bar\theta) \, {\tilde A}_y (X^{(1)}, Y^{(1)}) \Big]  \nonumber\\
&+& \Big[\partial_x  \left({\tilde A}_y (X^{(1)}, Y^{(1)})\right) 
- \partial_y \left( {\tilde A}_x (X^{(1)}, Y^{(1)}) \right) \Big]
{\bar\Psi}^{(1)} (t, \bar\theta)\,{\Psi}^{(1)}(t, \bar\theta) \nonumber\\
&+& i\, {\bar\Psi}^{(1)} (t, \bar\theta)\,{\dot{\Psi}}^{(1)}(t, \bar\theta),
\end{eqnarray}
where all the supervariables, present in the Lagrangian  ${\tilde L}^{(ac)}_0$, are the {\it ones} that 
have been derived  in (14) as well as (16) and the superscript ($ac$) stands for the anti-chiral behavior 
of the Lagrangian  ${\tilde L}^{(ac)}_0$. In view of the mapping (15), the invariance of the 
starting Lagrangian $L_0$ under $s_1$ can be captured within the framework of the supervariable approach as:   
\begin{eqnarray}
\frac{\partial}{\partial \bar\theta}\, {\tilde L}^{(ac)}_0 = - \;\frac{d}{d t} \;\Bigl [(A_x 
- i\; A_y) \;\psi  \Bigr ]   \Longleftrightarrow   s_1\, L_0 = - \;\frac{d}{d t} \;
\Bigl [(A_x - i\; A_y) \;\psi  \Bigr ].
\end{eqnarray}
The above equation encapsulates the geometrical meaning of the invariance of the starting Lagrangian 
$L_0$. This can be stated in the language of the translation along  the Grassmannian  direction 
$\bar\theta$. In fact, the above equation (26) demonstrates that the SUSY Lagrangian  
${\tilde L}^{(ac)}_0$ of the theory is a sum of composite supervariables such that its translation 
along the Grassmannian $\bar\theta$-direction produces  a total time derivative in the ordinary spacetime.

Exactly the above kind of analysis can be performed  for the invariance  of the starting 
Lagrangian  $L_0$ under the  SUSY transformations $s_2$. For instance, 
it can be checked that the starting Lagrangian $L_0$
can be generalized, onto the (1, 1)-dimensional  {\it chiral} super-submanifold,  as
\begin{eqnarray}
L_0 &\Rightarrow &     {\tilde L}^{(c)}_0 = \frac{1}{2}\,\Big[{\dot X}^{(2)}(t, \theta)
\,{\dot X}^{(2)}(t, \theta)  + {\dot Y}^{(2)}(t, \theta)\,{\dot Y}^{(2)}(t, \theta)\Big]
  \nonumber\\ 
&-& \Big[{\dot X}^{(2)}(t, \theta) \, {\tilde A}_x (X^{(2)}, Y^{(2)}) 
{\dot Y}^{(2)}(t, \theta) \, {\tilde A}_y (X^{(2)}, Y^{(2)}) \Big] \nonumber\\
&+& \Big[\partial_x  \left({\tilde A}_y (X^{(2)}, Y^{(2)})\right) 
- \partial_y  \left({\tilde A}_x (X^{(2)}, Y^{(2)}) \right) \Big]
{\bar\Psi}^{(2)} (t, \theta)\,{\Psi}^{(2)}(t, \theta) \nonumber\\
&+&  i\, {\bar \Psi}^{(2)} (t, \theta)\,{\dot{\Psi}}^{(2)}(t, \theta),
\end{eqnarray}
where all the supervariables of ${\tilde L}^{(c)}_0$ owe their origin to the superexpansions (22) and
(23) and the   superscript ($c$) on the Lagrangian shows its chiral behavior. In view of the 
relationship in (24), it is obvious that 
\begin{eqnarray}
\frac{\partial}{\partial \theta}\, {\tilde L}^{(c)}_0 =  \frac{d}{d t} \Big[\bar \psi \,
\bigl \{\dot x + i\, \dot y - (A_x + i\, A_y) \bigr \} \Big] \nonumber\\ \Longleftrightarrow   s_2\, L_0 = 
 \frac{d}{d t} \Big[\bar \psi \, \bigl \{\dot x + i\, \dot y - (A_x + i\, A_y) \bigr \} \Big].
\end{eqnarray}
The above relationship provides the geometrical meaning for the invariance  of  starting Lagrangian 
$L_0$ in the ordinary space under $s_2$. 

Now we concentrate on the geometrical interpretation for the nilpotency of the supercharges 
$Q(\bar Q)$ in the language of the translational generators $(\partial_\theta, \partial_{\bar\theta})$
along the ($\theta, \bar\theta$) directions   of the (1, 1)-dimensional super-submanifolds
of the general (1, 2)-dimensional supermanifold. Towards this goal in mind,  we note that we
can express the supercharge $Q$ in terms of the anti-chiral supervariables, 
in {\it three} different ways as: 
\begin{eqnarray}
Q &=&\frac{\partial}{\partial \bar\theta}\, \Big[- i\,\bar\Psi^{(1)}(t, 
\bar\theta)\,\Psi^{(1)}(t, \bar\theta)\Big] \nonumber\\ &\equiv &   \int d\bar\theta\,
\Big[- i\,\bar\Psi^{(1)}(t, \bar\theta)\,\Psi^{(1)}(t, \bar\theta)\Big],\nonumber\\ 
Q &=&\frac{\partial}{\partial \bar\theta}\, \Big[\Big(\dot x(t) 
- i\,\dot y(t)\Big)\, X^{(1)}(t, \bar\theta)\Big] \nonumber\\  &\equiv &  \int d\bar\theta\,
\Big[\Big(\dot x(t) - i\,\dot y(t)\Big)\, X^{(1)}(t, \bar\theta)\Big],\nonumber\\
Q &=&\frac{\partial}{\partial \bar\theta}\, \Big[i\,\Big(\dot x(t) 
- i\,\dot y(t)\Big) Y^{(1)}(t, \bar\theta)\Big] \nonumber\\ &\equiv&   \int d\bar\theta\,
\Big[i\,\Big(\dot x(t) - i\,\dot y(t)\Big) Y^{(1)}(t, \bar\theta)\Big],   
\end{eqnarray}
where the ordinary variables are from (1) and the supervariables are from
the super-expansions  (14) and (16).

In view of the mapping (15), the above charge ($Q$) can be {\it also}  expressed in 
the ordinary space as follows: 
\begin{eqnarray}
&&Q =s_1\, \Big[- i\,\bar\psi\,\psi\Big],  \;\; Q =s_1\, \Big[(\dot x -i \,\dot y)\,x\Big], 
\;\; Q =s_1\, \Big[i\,(\dot x -i \,\dot y)\, y \Big].  
\end{eqnarray}
Now the nilpotency of the charge $Q$ becomes pretty   trivial in the sense 
that it is connected with the nilpotency of the transformations $s_1$ through 
the relationship: $s_1\, Q = +\,i\,\{Q,\, Q\} = 0$ due to $s_1^2 = 0$. This observation could be 
also captured in the language of the translational generator $\partial_{\bar\theta}$ because  we 
observe that $\partial_{\bar\theta} \, Q = 0$ (due to expressions   of $Q$ listed in (29)) 
where we note that it is the nilpotency of the translational generator $\partial_{\bar\theta}$
(i.e. $\partial_{\bar\theta}^2 = 0$) which is responsible for the proof of the  nilpotency 
of $Q$. 

We focus on the nilpotency of $\bar Q$ in the language of geometry on the chiral super-submanifold. 
Towards this goal in mind,  
we can also express the supercharge $\bar Q$ in terms of the chiral supervariables, 
obtained after the application of SUSYIRs (20), in {\it three} different ways as: 
\begin{eqnarray}
\bar Q &=&\frac{\partial}{\partial \theta}\,
 \Big[ i\,\bar\Psi^{(2)}(t, \theta)\,\Psi^{(2)}(t, \theta)\Big] \nonumber\\
&\equiv &  \int d\theta\,\Big[i\,\bar\Psi^{(2)}(t, \theta)\,\Psi^{(2)}(t, \theta)\Big],\nonumber\\ 
\bar Q &=&\frac{\partial}{\partial \theta}\, \Big[X^{(2)}(t, \theta)\, \Big(\dot x(t) 
+  i\,\dot y(t)\Big)\Big]  \nonumber\\ &\equiv & \int d \theta\,
\Big[ X^{(2)}(t, \theta)\, \Big(\dot x(t) + i\,\dot y(t)\Big)\Big],\nonumber\\
\bar Q &=&\frac{\partial}{\partial \theta}\, \left[- i\,Y^{(2)}(t, \theta)\,\Big(\dot x(t) 
+ i\,\dot y(t)\Big)\, \right]  \nonumber\\ &\equiv &  \int d \theta\,
\left[- i\,Y^{(2)}(t, \theta) \Big(\dot x(t) + i\,\dot y(t)\Big)\right],   
\end{eqnarray}
where the ordinary variables are from (1) and the supervariables are from the expansions  (22) and (23).
In view of the mapping (24), we can express  (31) in the ordinary space
as: 
\begin{eqnarray}
&&\bar Q =s_2 \, \Big[ i\,\bar\psi\,\psi\Big],  \quad \bar Q =s_2\, \Big[x \,(\dot x +i \,\dot y)\Big], 
\quad \bar Q =s_2\, \Big[- i\, y \,(\dot x +i \,\dot y) \Big].  
\end{eqnarray}
The above two equations (31) and (32) show that the charge $\bar Q$ can be expressed in terms of 
 nilpotent ($s_2^2 = 0$) transformations $s_2$ and nilpotent ($\partial_\theta^2 = 0$) 
translational generator ($\partial_\theta$).

A close look at (31) and (32) clarify the nilpotency of the   charge $\bar Q$ which is beautifully intertwined 
with the nilpotency of $s_2$ (i.e. $s_2^2 = 0$) and/or nilpotency ($\partial_\theta^2 = 0$)
of the translational generator $\partial_\theta$ on the chiral super-submanifold.
This can be verified  by the observation that $s_2\,\bar Q = +\,i\,\{\bar Q,\,\bar Q\} = 0$
due to nilpotency of $s_2$. Similarly, we note that $\partial_\theta\, \bar Q = 0$ because of the nilpotency 
($\partial_\theta^2 = 0$) of the generator $\partial_\theta$.

\section{Cohomological aspects: Continuous $\mathcal{N} = 2$ SUSY symmetries}

For the sake of completeness of our paper, we concisely  point out the mathematical meaning of the 
symmetry transformation operators ($s_1, s_2, s_\omega$) that have been mentioned in equations (2) and (4).
Towards this goal in mind, we modify the transformations (2) by a constant 
factor\footnote{We have taken a factor of ($1/{\surd 2}$) in the overall transformations 
so that the corresponding charges would be able to satisfy one of the simplest form of the $sl(1/1)$ 
algebra of $\mathcal {N} = 2$ SUSY quantum mechanics (cf. (40) below) where there is no central extension.}  
as  [17] 
\begin{eqnarray}
&& s_1 x = \frac{\psi}{\surd 2}, \qquad s_1 y = \frac{-i\;\psi}{\surd 2}, \qquad
 s_1 \bar \psi = \frac{i}{\surd 2}\, [\dot x - i \;\dot y ], \qquad s_1 \psi = 0, \nonumber\\ &&
 s_1 A_x = \frac{1}{\surd 2}\, \bigl (\partial_x A_x - i\; \partial_y A_x \bigr ) \, \psi, \;\;\qquad
s_1 A_y = \frac{1}{\surd 2}\, \bigl (\partial_x A_y - i \;\partial_y A_y \bigr )\, \psi, \nonumber\\ 
&&  s_2 x = \frac{\bar \psi}{\surd 2}, \qquad  
s_2 y = \frac{i\;\bar \psi}{\surd 2}, \qquad   
s_2  \psi = \frac{i}{\surd 2}\, [\dot x + i \;\dot y ],  \qquad s_2 \bar \psi = 0, \nonumber\\
&& s_2 A_x = \frac{\bar \psi}{\surd 2}\, \bigl (\partial_x A_x 
+ i \;\partial_y A_x \bigr ),\qquad 
s_2 A_y = \frac{\bar \psi}{\surd 2}\, \bigl (\partial_x A_y + i \;\partial_y A_y \bigr ). 
\end{eqnarray}
It is straightforward to check that the algebra
obeyed by the transformation operators ($s_1, s_2, s_\omega$) is\footnote{It is 
elementary to note that the  transformations (4) (i.e. $s_\omega$) would be now 
expressed modulo an $i$ factor because of the modifications in (33) {\it vis-\`a-vis} (2).}:
\begin{eqnarray}
&& s^2_1  = 0,\,\quad s^2_2  = 0, \,\quad \{s_1,\, s_2\} = s_\omega = (s_1 + s_2)^2, \nonumber\\
&& \big[s_\omega,\, s_1 \big] = 0,\qquad [ s_\omega,\, s_2 ] = 0,\qquad \{s_1,\,s_2\} \ne 0.
\end{eqnarray}
At the algebraic level, the above algebra  is exactly  like the algebra obeyed by the de Rham cohomological 
operators (see, e.g. [18-21]
\begin{eqnarray}
&&d^2 = 0,\qquad \delta^2 = 0, \qquad \{d,\, \delta\} = \Delta = (d + \delta)^2,\nonumber\\
&&\big[\Delta,\, d \big] = 0, \;\;\qquad \big[\Delta,\, \delta \big] = 0, \;\;\qquad \{d,\, \delta\} \ne 0. 
\end{eqnarray} 
In the above, the operators  $(\delta)d$ are the (co-)exterior derivatives (with $d^2 = \delta^2 = 0$) and 
$\Delta = (d + \delta)^2$ is the absolutely {\it commuting} Laplacian  operator.

In the realm  of differential geometry, one knows that the (co-)exterior derivatives are connected by the 
relation $\delta =  \pm\,* \,d\, *$ where ($*$) is the Hodge duality operation on a given compact 
manifold on which the set ($d,\, \delta,\, \Delta$) is defined. In our theory, the ($*$)
operation is replaced by a discrete set of {\it symmetry} transformations:
\begin{eqnarray}
&& x \to  \mp \;x,  \quad \psi \to \mp \;\bar \psi, \quad A_x \to \pm\; A_x, \quad t \to -\; t, \nonumber\\
&& y \to \pm\; y, \quad \bar\psi \to \pm  \;\psi, \quad A_y \to \mp\; A_y, \quad B_z \to B_z,
\end{eqnarray} 
under which the Lagrangian  (1) remains invariant and  we observe that the 
nilpotent transformations $s_2$ and $s_1$ are 
connected by  [17]
\begin{eqnarray}
&& s_2 \Phi_1 = +  * s_1\, *\, \Phi_1 \quad \,\Rightarrow s_2 = + *\, s_1\, *,\quad \qquad
\Phi_1 = x, y, A_x, A_y,   \nonumber\\
&& s_2 \Phi_2 = -\; * s_1\; *\; \Phi_2 \;\;\Rightarrow\;\; s_2 = - \;*\; s_1\; *, \qquad
 \Phi_2 = \psi, \bar\psi,
\end{eqnarray} 
where  ($\pm$) signs, in  the above relationship, are governed by the application of
two consecutive  discrete symmetry transformations, namely; 
\begin{eqnarray}
&& *\; [\; *\; ] \; \Phi_1 = \; +\; \Phi_1,  \qquad \qquad \Phi_1 = x, y, A_x, A_y, \nonumber\\
&& *\; [\; *\; ] \; \Phi_2 = \; -\; \Phi_2,  \qquad \qquad \Phi_2 = \; \psi , \; \bar \psi.
\end{eqnarray}
The above is the rule (for signatures in (37)) laid down by the requirements 
of a perfect {\it duality} invariant theory
[22]. We note, from equation (38), that it is the interplay of continuous
 and discrete symmetry transformations 
of our theory which provide the physical realization of the relationship between the (co-)exterior 
derivatives: $\delta = \pm\, *\,d\,*$ of differential geometry.

For our model under consideration,  we note that 
the continuous symmetry transformations ($s_1, s_2, s_\omega$)  
lead to the following expressions for the  Noether conserved charges $Q_i$ (with $i = 1, 2, 3)$, namely;
\begin{eqnarray}
 Q_1 &\equiv & Q = \frac{1}{2} \,\Bigl [ (p_x + A_x) - i \;(p_y + A_y) \Bigr ] \, \psi, \nonumber\\
 Q_2 &\equiv & \bar Q =  \frac{\bar\psi}{2} \, \Bigl [ (p_x + A_x) + i \;(p_y + A_y) \Bigr ], \nonumber\\
  Q_3 & \equiv &   Q_\omega =  \Big[\frac{(p_x + A_x)^2}{2} +  \frac{(p_y + A_y)^2}{2}  
- B_z\; \bar\psi\; \psi \Big] \equiv  H.
\end{eqnarray}
It can be readily checked  that the above charges obey one of the simplest $\mathcal{N} = 2$
SUSY quantum mechanical algebra, namely; 
\begin{eqnarray}
&& Q^2 = 0 \qquad\qquad \bar Q^2 = 0, \qquad \qquad   \{Q, \;\bar Q \} = H, \nonumber\\
&& \dot Q = - \,i\, [Q, \;H] = 0,\;\;\;\;\qquad  \dot{\bar Q} = - \,i\,[\bar Q, \, H] = 0,
\end{eqnarray}
which provides the physical realization of the Hodge algebra. 

\section{Conclusions}

In our present endeavor, we have taken an example of the $\mathcal{N} = 2$ SUSY quantum mechanical model 
whose superpotential is totally different from the cases of $\mathcal{N} = 2$ SUSY free particle  
and HO (and the generalization of HO) [14,15,17]. This has been done 
{\it purposely} so that our idea of the supervariable
approach [14,15] could be put on a solid foundation. We have derived the proper $\mathcal{N} = 2$ transformations 
for the SUSY system under consideration by exploiting the idea of SUSYIRs. We have also 
provided the geometrical  basis for the nilpotency of SUSY transformations 
and SUSY invariance of the Lagrangian in the language of translational generators 
($\partial_\theta,\, \partial_{\bar\theta}$) on the (1, 1)-dimensional  
chiral and anti-chiral super-submanifolds.

We have demonstrated, in our present investigation, that the nilpotency of a SUSY transformation
of an ordinary dynamical variable (of the starting Lagrangian (1)) is intimately connected with a 
set of two successive translations of the corresponding supervariable along 
$(\bar\theta)\theta$ directions of the (1, 1)-dimensional
(anti-)chiral super-submanifolds of the general (1, 2)-dimensional supermanifold on which our starting
theory is generalized (cf. Sec. 4). Similarly, we have established that the SUSY invariance of the
Lagrangian (1) is equivalent to the translation of a sum of composite supervariables (that are present in the
(anti-)chiral Lagrangians) along the  Grassmannian $(\bar\theta)\theta$ 
directions of the (anti-)chiral super-submanifolds such that this process yields  a total 
time derivative in the ordinary space.

Our present work and earlier works [14,15] are our modest {\it first} few steps towards our main goal 
of deriving the SUSY transformations with the minimal knowledge  about the {\it classical} 
Lagrangian and its symmetries. 
Such expectations and intuitions   have been spurred  due to our experiences in the 
application of superfield formalism [4,5,9-13]
to the gauge systems.
In fact, in the realm  of BRST formalism, if one knows the (anti-)BRST symmetries, 
there is absolutely  {\it no} problem in obtaining the gauge-fixing and Faddeev-Popov
ghost terms (see, e.g. [9-13]). Our central  ideology is to develop theoretical 
tools and techniques so that we could derive the whole structure of the SUSY invariant Lagrangian from 
the knowledge of SUSY symmetry transformations that emerge from the supervariable approach. 

So far, we have applied our supervariable approach to the derivation of $\mathcal{N} = 2$
SUSY symmetries  for some explicit examples, viz., $\mathcal{N} = 2$ SUSY free particle and HO.
Our main goal is to apply the augmented  version of superfield approach [9-13] to the SUSY gauge theories 
that have become important  because of their relevance to the modern developments in superstring  
theories. In fact, our aim is to study the $\mathcal{N} = 2,\, 4$ and $8$ SUSY
gauge theories within the framework  of BRST formalism where, we are sure, our augmented version of BT-superfield 
formalism [9-13] would play very important role. As far as SUSY gauge  theories are concerned,
we have already taken the {\it first} modest step and
supersymmetrized the HC in the context of  (SUSY system of) 
a free spinning relativistic particle and obtained 
the proper (i.e. nilpotent and absolutely anticommuting) 
(anti-)BRST transformations [23]. Presently, one of us is also involved with the application of 
super-HC to derive the (anti-)BRST symmetries in the context of Abelian SUSY gauge theory [24].\\

\noindent
{\bf Acknowledgements:} \\

\noindent
One of us (SK) would like to gratefully acknowledge financial support from UGC (RGNF), Govt. of
India, New Delhi, under grant No. F1-17.1/2011-12/RGNF-SC-UTT-708/(SA-III/Website) of the SRF-scheme.
We thank the Reviewer for very useful comments.\\

\noindent
{\bf{\large{Appendix A: Choice of the (anti-)chiral supervariables for  the 
  $~~~~~~~~~~~~~~~~~~~$ description  of our SUSY model}}}\\ 

\noindent
As we have mentioned in the main body of our text, one of the key differences  between  the $\mathcal{N} = 2$
SUSY transformations and (anti-)BRST symmetry transformations is the fact that whereas the latter are absolutely 
anticommuting, the former are {\it not}. Thus,  within  the framework of BT-superfield approach to
(anti-)BRST symmetry transformations, a generic superfield (defined on a (D, 2)-dimensional
supermanifold) is expanded  along both the Grassmannian directions ($\theta$ and $\bar\theta$)
of the supermanifold, namely;
\[\Sigma  (x, \theta, \bar\theta) =  \sigma (x) + \theta \, \bar R(x)
 +\bar\theta\, R(x) + i\;\theta\;\bar\theta\, S(x),\eqno(A.1)\] 
where $\sigma (x)$ is an ordinary $D$-dimensional field of the original (anti-)BRST invariant theory 
and $\Sigma  (x, \theta, \bar\theta)$ is the corresponding  superfield on the 
(D, 2)-dimensional supermanifold that is  characterized  by the superspace 
coordinates $Z^M = (x^\mu,\theta,\bar\theta)$ (with $\theta^2 = {\bar\theta}^2 = 0,\, \theta\, \bar\theta 
+ \bar\theta\, \theta = 0$).

It is evident, from (A.1), that {\it if} $\sigma(x)$ is a bosonic ordinary field, 
then, $\Sigma  (x, \theta, \bar\theta)$ would be also bosonic 
(i.e. secondary  fields ($R,\, \bar R$) would be fermionic and $S(x)$ bosonic).
On the contrary, if $\sigma(x)$ is fermionic, then, $\Sigma  (x, \theta, \bar\theta)$ and $S(x)$ 
would be fermionic, too. The pair $(R, \bar R)$  would become  bosonic in the case of $\sigma(x)$
being fermionic. A natural  consequence of the expansion in (A.1) is the observation that
\[\frac{\partial}{\partial{\bar\theta}} \,\frac{\partial}{ \partial{\theta}}
\,\Big( \Sigma (x, \theta, \bar\theta)\Big) 
= i\, S(x)\quad \Longleftrightarrow \quad s_b \,s_{ab}\, \sigma (x),\] 
\[ \frac{\partial}{\partial{\theta}} \, \frac{\partial}{\partial {\bar\theta}}\,
\Big( \Sigma (x, \theta, \bar\theta)\Big) 
=- i\, S(x) \; \Longleftrightarrow \; s_{ab} \,s_{b}\, \sigma (x), \eqno(A.2)\]
where $s_{(a)b}$  are the (anti-)BRST symmetries and they are identified  with the translational
generators ($\partial_\theta)\partial_{\bar\theta}$ along the Grassmannian direction $(\theta)\bar\theta$ 
of the (D, 2)-dimensional supermanifold [4-13].
It is clear, from (A.2), that  
\[(\partial_{\bar\theta} \, \partial_{\theta}
+ \partial_{\theta} \, \partial_{\bar\theta}) \, \Sigma (x, \theta, \bar\theta) = 0
\;\;\Longleftrightarrow   \quad(s_b\, s_{ab} + s_{ab}\, s_b)\,\sigma (x)  = 0, \eqno(A.3)\]
which establishes the absolute anticommutativity of the (anti-)BRST symmetry transformations.
Furthermore, it is also obvious that the nilpotency (i.e. $\partial_\theta^2 = \partial_{\bar\theta}^2 = 0$)
of the above translational generators ($\partial_\theta)\partial_{\bar\theta}$  implies
the off-shell nilpotency of the fermionic (anti-)BRST transformations (i.e. $s^2_{(a)b} = 0$). 
Thus, whenever we consider the full expansions 
(like (A.1)) of the superfield, the nilpotency and absolute anticommutativity properties are
 automatically implied within the framework of superfield formalism (see, e.g. [4-13]).

 The application of our supervariable approach to a SUSY system theoretically compels   us to choose the 
 (anti-)chiral supervariables so that we could capture {\it only} the nilpotency property but {\it avoid}
 the absolute anticommutativity of the $\mathcal{N} = 2$ SUSY transformations. 
 The latter property is a decisive  feature of the $\mathcal{N} = 2$ SUSY quantum mechanical theory. 

\end{document}